\documentclass[aps,prl,twocolumn,preprintnumbers,%
               floatfix,nofootinbib]{revtex4}
\usepackage{graphicx}

\def\beq{\begin{equation}}
\def\eeq#1{\label{#1}\end{equation}}
\def\eeqn{\end{equation}}
\def\beqa{\begin{eqnarray}}
\def\eeqa#1{\label{#1}\end{eqnarray}}
\def\eeqan{\end{eqnarray}}
\def\CR{\nonumber \\ }
\def\leqn#1{(\ref{#1})}
\def\({\left(}
\def\){\right)}

\def\stacksymbols #1#2#3#4{\def\theguybelow{#2}
    \def\vp{\lower#3pt}
    \def\sp{\baselineskip0pt\lineskip#4pt}
    \mathrel{\mathpalette\intermediary#1}}

\def\intermediary#1#2{\vp\vbox{\sp
     \everycr={}\tabskip0pt
     \halign{$\mathsurround0pt#1\hfil##\hfil$\crcr#2\crcr
              \theguybelow\crcr}}}

\def\to{\rightarrow}

\begin{document}

%\wideabs{
%\begin{flushright}
%{\tt hep-ph/0403004} \\
%UFIFT-HEP-04-3 \\
%\end{flushright}

\title{Dark Matter at Colliders: a Model-Independent Approach}   
\author{Andreas Birkedal$^{1,2}$, Konstantin Matchev$^{1,2}$, and Maxim 
Perelstein$^1$}
\address{$^1$ Institute for High-Energy Phenomenology,
Cornell University, Ithaca, NY~14853 \\
$^2$ Physics Department, University of Florida, Gainesville, FL 32611}
%\date{\today}
%\maketitle

\begin{abstract}
\noindent
Assuming that cosmological dark matter consists of weakly interacting
massive particles, we use the recent precise measurement of cosmological 
parameters to predict the guaranteed rates of production of such particles 
in association with photons at electron-positron colliders. Our approach is 
based on general physical principles such as detailed balancing and 
soft/collinear factorization. It leads to predictions that are valid across 
a broad range of models containing WIMPs, including supersymmetry, universal 
extra dimensions, and many others. We also discuss the discovery prospects 
for the predicted experimental signatures.
\end{abstract}

\maketitle
%}

{\it Introduction ---} A variety of astrophysical and cosmological 
observations indicate that a substantial fraction (perhaps as much as
30\%) of the energy density in the Universe is due to non-relativistic,
non-baryonic, non-luminous matter. The microscopic nature of this ``dark''
matter is at present unknown. Perhaps the most attractive
explanation is provided by the {\em ``WIMP hypothesis''}: dark matter is
assumed to consist of hypothetical stable particles with masses around the 
scale of electroweak symmetry breaking, in the 10 GeV -- 1 TeV range, whose 
interactions with other elementary particles are of the strength and range
similar to the familiar weak interactions of the Standard Model. Such weakly 
interacting massive particles (WIMPs) naturally have a relic abundance of 
the correct order of magnitude to account for the observed dark 
matter, making them appealing from a
theoretical point of view. Moreover, many extensions of the Standard Model
contain particles which can be identified as WIMP dark matter candidates.
Examples include supersymmetric models~\cite{SUSY}, models with ``universal'' 
extra dimensions~\cite{UED}, little Higgs theories~\cite{LH}, etc. 

Particle accelerators of the next generation, such as the Large Hadron 
Collider (LHC) at CERN and the proposed linear electron-positron collider, 
may have enough energy to produce WIMPs. Once produced, WIMPs escape the 
detector without interactions, leading to an apparent 
energy imbalance, or ``missing energy'' signature. In this letter, we will
use the known cosmological abundance of WIMPs to predict the rate of such 
events at future colliders. In sharp contrast to all existing studies, we
will do so without making {\it any} assumptions about the details of the
particle physics model responsible for the WIMP: our results are equally 
valid in all theories listed in the previous paragraph, as well as in any
other model containing WIMPs. In our approach, all 
model uncertainties are reduced to a single continuous parameter 
with a transparent physical meaning, plus a  
small number of discrete choices such as the spin of the WIMP.

{\em WIMP Annihilation Cross Sections from Cosmology ---} We assume that the 
observed dark matter is entirely due
to a single WIMP particle, $\chi$. This particle carries a new conserved 
quantum number which prevents it from decaying into lighter Standard Model 
particles, making it stable. At the same time, two WIMPs can annihilate
into a pair of Standard Model particles~\cite{footnote1}:
\beq
\chi+\chi \to X_i + \bar{X}_i,
\eeq{annih} 
where $X_i=l,q,g,\ldots$ can be any Standard Model particle.  
We assume that~\leqn{annih} is the only process important for the 
determination of the $\chi$ relic abundance; that is, we ignore 
the possibility of coannihilations between $\chi$'s and other 
exotic particles.
%affect this determination will be considered in a subsequent 
%publication~\cite{future}. 
The present dark matter density depends on the 
cross sections of reactions~\leqn{annih} in the limit when the colliding 
$\chi$ particles are non-relativistic. If $v$ is the relative velocity of 
two $\chi$'s, each cross section can be expanded as~\cite{but}
\beq
\sigma_i v = \sum_{J=0}^{\infty} \sigma_i^{(J)} v^{2J},
\eeq{expand}
where $\sigma^{(0)}$ only receives a contribution from $s$-wave 
annihilation, $\sigma^{(1)}$ receives contributions from $s$- and $p$-wave
channels, etc. It is clear that at low $v$, 
the lowest non-vanishing term in Eq.~\leqn{expand} will dominate. We define
\beq
\sigma_{\rm an} = \sum_i \sigma_i^{(J_0)},  
\eeq{total}
where $J_0$ is the angular momentum of the dominant partial wave contributing 
to %any 
$\chi$ annihilation %process 
in a given model, and the sum only runs 
over the final states that give a non-vanishing contribution at this order 
in $J$. For most of this letter, we will restrict our analysis to 
two cases: $J_0=0$ and $J_0=1$. We will refer to WIMPs in each case as
``$s$-annihilators'' and ``$p$-annihilators'', respectively. (The extension 
of our analysis to higher $J_0$ is trivial but rather unmotivated since all 
known models predict either $s$-wave or $p$-wave annihilation.)

\begin{figure}[tb]
\begin{center}
\includegraphics[scale=0.45]{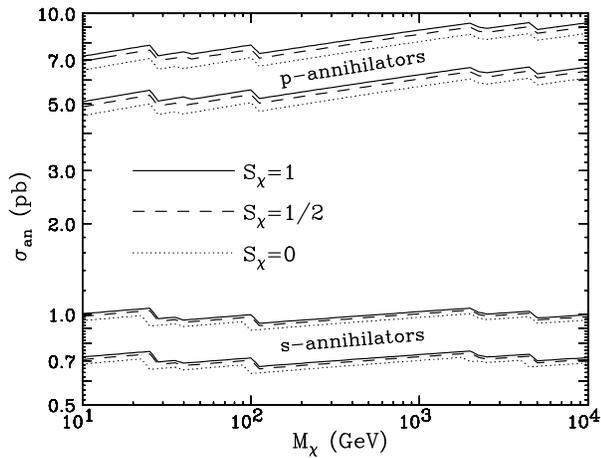}
\vskip5mm
\caption{Values of the quantity $\sigma_{\rm an}$ defined in~\leqn{total}
allowed at $2\sigma$ level as a function of the WIMP mass. The lower (upper)
band is for models where $s$-wave ($p$-wave) annihilation dominates.}
\label{fig:WMAP}
\end{center}
\end{figure}

The present cosmological abundance of WIMPs is mainly determined by 
the values of $J_0$ and $\sigma_{\rm an}$, with only a weak dependence on
other parameters such as the WIMP mass $M_\chi$, its spin $S_\chi$, etc.  
It is therefore these parameters that are strongly constrained by cosmological 
observations. In Figure~\ref{fig:WMAP}, we show the cosmological constraint 
on $\sigma_{\rm an}$, as a function of $M_\chi$, 
\noindent for $s$- and
$p$-annihilators and three values of the WIMP spin, $S_\chi=0, 1/2, 1$. We 
have used the WMAP value of the present dark matter abundance,
$\Omega_{\rm dm}h^2=0.112\pm 0.009$~\cite{DMref}. The constraint on 
$\sigma_{\rm an}$ is surprisingly robust, as the logarithmic dependence on
$M_\chi$ is offset by the change in the effective number of degrees of
freedom at different freeze-out temperatures.

{\em Detailed Balancing and WIMP Production at Colliders~---} 
%Assuming that $CP$ violating couplings do not give significant contributions 
%to the WIMP annihilation processes, 
The cross sections of the reaction~\leqn{annih} and its inverse 
are related by the detailed balancing equation~\cite{Frazer}:
\beq
\frac{\sigma(\chi+\chi\to X_i + \bar{X}_i)}
{\sigma(X_i+\bar{X}_i\to \chi+\chi)}\,=\,
2\,\frac{v_{\rm X}^2 (2S_{\rm X}+1)^2}{v_\chi^2 (2S_\chi+1)^2}\,,
\eeq{detailed}
where the cross sections are averaged over spins but not other quantum numbers 
such as color. For each Standard Model particle $X_i$, we define the
``annihilation fraction'' $\kappa_i$ as
\beq
\kappa_i = \frac{\sigma_i^{(J_0)}}{\sigma_{\rm an}}.
\eeq{kappa_def}   
Note that $\sum\kappa_i=1$.
Using Eqs.~\leqn{detailed} and~\leqn{kappa_def}, we obtain the following 
expression for the production of non-relativistic $\chi$ pairs in 
$X_i\bar{X_i}$ collisions
\beqa
& &\sigma(X_i\bar{X}_i\to2\chi) = \CR & &
2^{2(J_0-1)}\,\kappa_i \sigma_{\rm an}\,
\frac{(2S_\chi+1)^2}{(2S_{\rm X}+1)^2}   
\left(1-\frac{4M_\chi^2}{s}\right)^{1/2+J_0},
\eeqa{prediction} 
where we have assumed that the inital state particles are relativistic 
($M_X\ll M_\chi$). This formula is only valid at center of mass energies 
slightly above the $2\chi$ threshold, $v=2v_\chi=2\sqrt{1-4M_\chi^2/s}\ll 1$, 
and 
receives corrections of order $v^2$. Taking $X_i=q$ or $g$ (or even $W$, $Z$)
for a hadron collider or $X_i=e$ for an electron-positron machine, 
Eq.~\leqn{prediction} 
provides a prediction of the WIMP production rate. The model-dependence of 
this prediction is contained in a small number of parameters with a clear 
physical meaning: the mass $M_\chi$ and the spin $S_\chi$ of the WIMP, the 
value of $J_0$, and 
the annihilation fraction $\kappa_i$ for the given initial state. Crucially,
the overall scale for this prediction, the quantity $\sigma_{\rm an}$, is
provided by cosmology, as shown in Fig.~\ref{fig:WMAP}.  

{\em Tagging and Factorization ---} Unfortunately, the $2\chi$ production
process whose cross section we predicted is not measurable at 
colliders. Much like Standard Model neutrinos, WIMPs cannot be 
directly observed due to the weakness of their interactions with matter. At
least one detectable particle is required for the event to pass the triggers 
and be recorded on tape. 
%We thus need to consider processes with at least one 
%additional particle in the final state. 
In order to retain the model-independence of our analysis, we need to study 
processes in which two WIMPs are produced in association with a photon or 
a gluon radiated from the known initial state.

In this letter, we concentrate on the case of $e^+e^-\to2\chi+\gamma$. For 
general 
kinematics, there is no model-independent relation between the rate of this 
process and that of $e^+e^-\to2\chi$ predicted by Eq.~\leqn{prediction}. 
However, if the emitted photon is either {\it soft} or {\it collinear} with 
the incoming electron or positron, soft/collinear factorization theorems 
provide such a relation. Emission of collinear photons is described by 
\beq
\frac{d\sigma(e^+e^-\to 2\chi+\gamma)}{dx\, d\cos\theta} \approx
{\cal F}(x, \cos\theta)\,\hat{\sigma}(e^+e^-\to2\chi),
\eeq{collinear}
where $x=2E_\gamma/\sqrt{s}$ ($E_\gamma$ is the photon energy), 
$\theta$ is the angle between the photon direction and the direction of the 
incoming electron beam, ${\cal F}$ denotes the collinear factor:
\beq
{\cal F}(x, \cos\theta) = \frac{\alpha}{\pi}\frac{1+(1-x)^2}{x} 
\frac{1}{\sin^2\theta}\,,
\eeq{llog}
and $\hat{\sigma}$ is the WIMP pair-production cross section evaluated at 
the reduced center of mass energy, $\hat{s}=(1-x)s$. Note that upon
integration over $\theta$, the above equation reproduces 
the familiar Weizsacker-Williams distribution function. The factor ${\cal F}$ 
is universal: it does not depend on the nature of the (electrically neutral) 
particles produced in association 
with the photon. Emission of soft photons is described by the leading piece
of Eq.~\leqn{collinear} in the $x\to 0$ limit.

Combining Eqs.~\leqn{prediction} and \leqn{collinear}, we find
\beqa
& &\hskip-.8cm
\frac{d \sigma}{dx\, d\cos\theta} (e^+e^-\to 2\chi+\gamma)\,\approx\,\CR 
& &\hskip-.8cm 
\frac{\alpha\kappa_e \sigma_{\rm an}}{16\pi} \frac{1+(1-x)^2}{x} 
\frac{1}{\sin^2\theta}2^{2J_0} (2S_\chi+1)^2\,  
\left(1-\frac{4M_\chi^2}{(1-x)s}\right)^{1/2+J_0}.
\eeqa{rate}
This formula, applicable for collinear photons ($\theta\to 0$ or 
$\theta\to\pi$), is the main result of this letter.

\begin{figure}[t!]
\begin{center}
\includegraphics[scale=0.23]{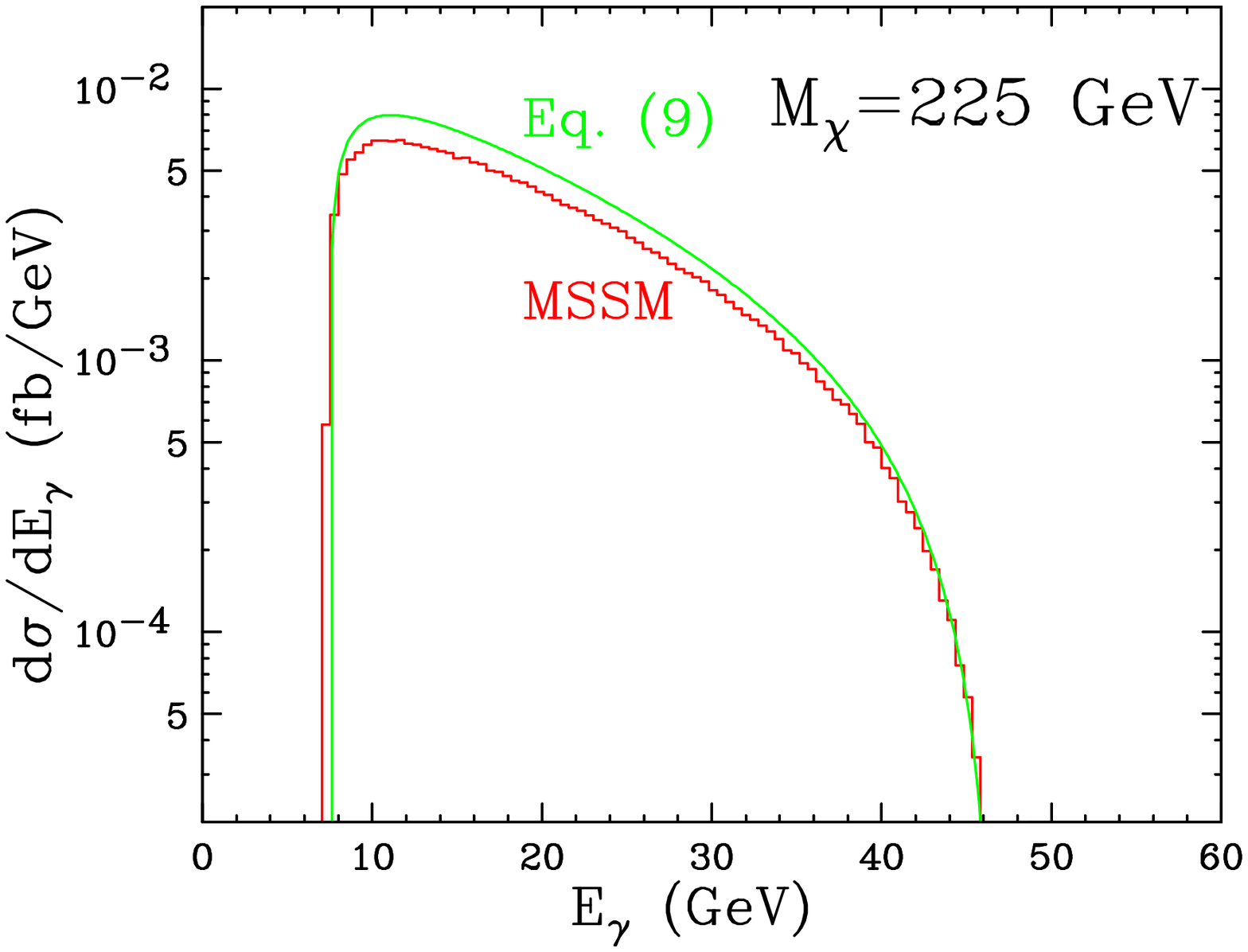}
\includegraphics[scale=0.23]{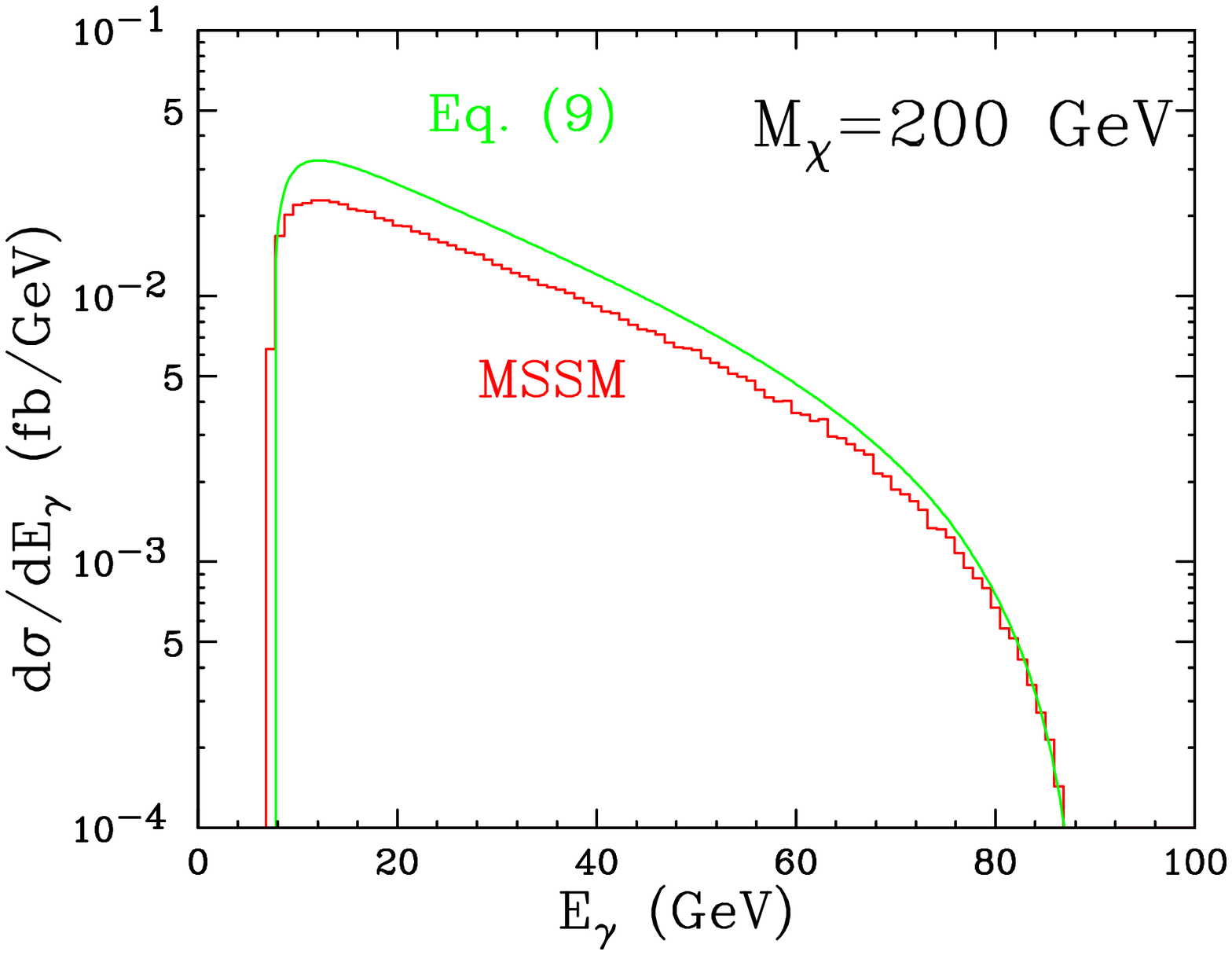}
\vskip5mm
\includegraphics[scale=0.23]{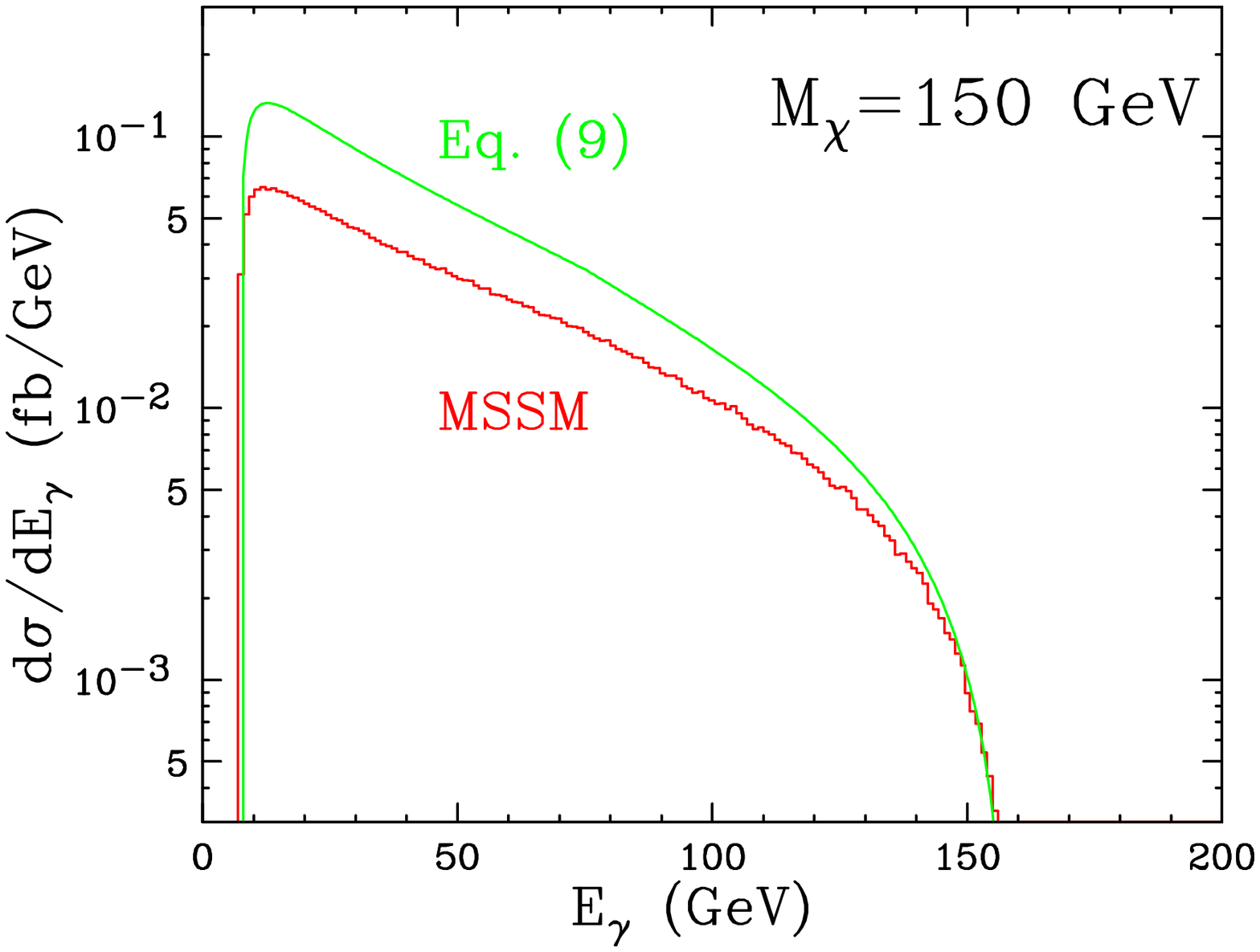}
\includegraphics[scale=0.23]{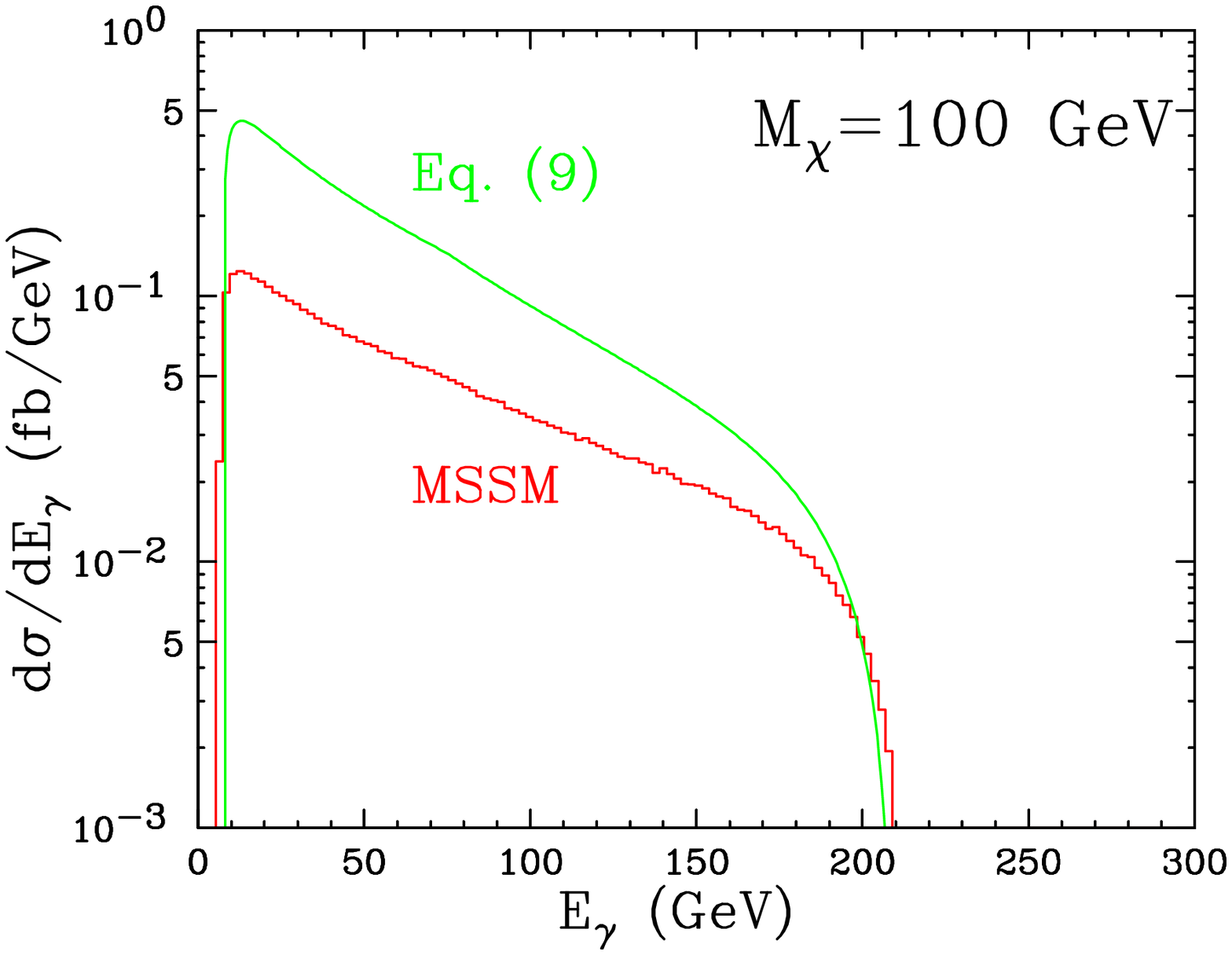}
\vskip5mm
\caption{Comparison between the photon spectra from the process 
$e^+e^-\to2\chi_1^0+\gamma$ in the explicit supersymmetric models defined in 
the text (red/dark-gray) and the spectra predicted by Eq.~\leqn{rate} for a 
$p$-annihilator of the corresponding mass and $\kappa_e$ 
(green/light-gray).}
\label{fig:comp1}
\end{center}
\end{figure}

{\it Validity of Collinear Approximation ---} Eq.~\leqn{rate}
predicts the rate of events with a single collinear photon and missing energy  
due to WIMP production in $e^+e^-$ collisions. However, very collinear photons 
cannot be detected in an experiment, due to an incomplete electromagnetic 
calorimeter coverage around the beam pipe as well as the lower cut on the 
$p_{T,\gamma}\equiv E_\gamma \sin\theta$ that has to be imposed to reject 
backgrounds such as $e^+e^-\to e^+e^-\gamma$ where electron and positron are 
too forward to be detected. Do
the predictions made using the collinear factorization approach have any value 
in realistic circumstances? To address this question, we have compared
the event rates obtained by integrating Eq.~\leqn{rate} with realistic cuts 
with those obtained in explicit models containing WIMPs without making any
approximations. For this comparison, we have chosen $\sqrt{s}=500$ GeV, 
assumed the electromagnetic calorimeter acceptance 
$\sin\theta>0.1$~\cite{Tesla}, and required $p_{T,\gamma}>7.5$ GeV 
corresponding to the mask calorimeter acceptance of 1 degree. The 
results of the comparison are shown in Figure~\ref{fig:comp1}. The red 
(dark-gray) histograms show the photon spectra from the 
reaction $e^+e^-\to \chi_1^0\chi_1^0 \gamma$ within the minimal supersymmetric 
standard model (MSSM)~\cite{CompHEP} with the parameters suitably chosen to
provide the correct neutralino relic density~\cite{darkSUSY}. (Explicitly, for 
$M_\chi=(100,150,200,225)$ GeV, the MSSM parameters take the following values 
at the weak scale: $M_1=(115,168,217,242)$ 
GeV, $\mu=(185,225,275,300)$ GeV, and $m_{\tilde{\ell}_R}=(115,177,237,268)$ 
GeV; for all four points, $M_2=2M_1$, $\tan\beta=10$, and all the mass 
parameters not listed above are fixed at 1 TeV.) The green (light-gray) 
lines on the same figure show the spectra predicted by Eq.~\leqn{rate} for 
a ``generic'' $p$-annihilator of the corresponding mass and $\kappa_e$.
We conclude that our approach works quite well. The photon 
spectrum near the endpoint is correctly reproduced for all $M_\chi$.
Eq.~\leqn{rate} fails for lower values of $E_\gamma$; this 
effect is especially noticeable for low $M_\chi$. This is due not to the 
failure of collinear approximation, but rather to the fact that the relative 
motion of the produced $\chi$ particles becomes relativistic in this regime, 
and the higher-order terms in the $v^2$ expansion of Eq.~\leqn{expand}, not 
captured by $\sigma_{\rm an}$, are important. 
Model-independent WIMP searches at $e^+e^-$ colliders, which we discuss 
below, should take this limitation into account by concentrating on the 
photons near the endpoint of the spectrum. Note that we did not have to 
impose an additional cut to eliminate central photons: collinear
emission naturally dominates the signal. 

%\begin{table}
%\begin{center}
%\begin{tabular}{|c||c|c|c|c|} \hline
%   $M_\chi$ & $M_1$ & $M_2$ & $\mu$ & $m_{\tilde{l}_R}$ \\ \hline
%100 & 115 & 230 & 185 & 123  \\ \hline
%150 & 168 & 336 & 225 & 182  \\ \hline
%200 & 217 & 434 & 275 & 240  \\ \hline
%225 & 242 & 484 & 300 & 270  \\ \hline
%\end{tabular}
%\caption{Weak-scale MSSM mass parameters (in GeV) used to predict the photon
%spectra shown in Figure~\ref{fig:comp1}. All the mass parameters not listed 
%here are fixed at 1 TeV, and $\tan\beta=10$ for all four panels.}
%\label{tab:MSSM}
%\end{center}
%\end{table}

\begin{figure}[tb]
\begin{center}
\includegraphics[scale=0.45]{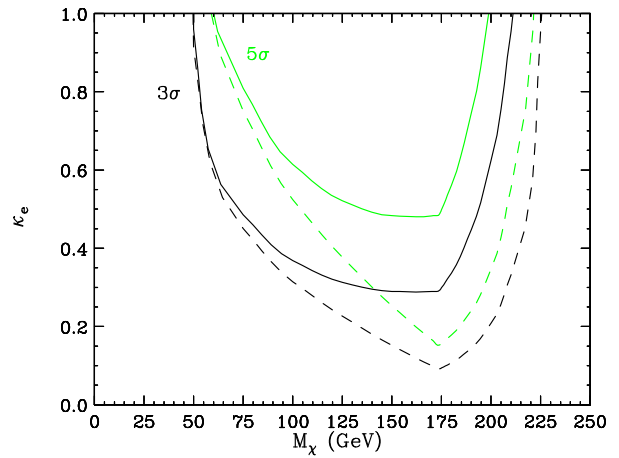}
\vskip5mm
\caption{The reach of a 500 GeV unpolarized electron-positron collider with 
an integrated luminosity of 500 fb$^{-1}$ for the discovery of 
$p$-annihilator WIMPs, as a function of the WIMP mass $M_\chi$ and the 
$e^+e^-$ annihilation fraction $\kappa_e$. The 3 $\sigma$ (black) and 5 
$\sigma$ (green/light-gray) contours are shown. The dashed lines include 
only statistical uncertainty, whereas the solid lines include a systematic 
uncertainty of 0.3\%~\cite{Tesla}.}
\label{fig:carrot}
\end{center}
\end{figure}

{\em Experimental Searches for WIMPs ---} The main irreducible background
to the search for anomalous $\gamma$+missing $E$ events is provided by the 
Standard Model reaction $e^+e^-\to\nu\bar{\nu}\gamma$. At the energies well
above the $Z$ peak, 
this reaction is dominated by the $t$-channel $W$ exchange contribution, and 
has a rather large cross section. Nevertheless, the enhancement of the rate 
predicted by Eq.~\leqn{rate} may well be observable. In 
Figure~\ref{fig:carrot}, we show the reach of a 500 GeV linear collider (LC) 
with an integrated 
luminosity of 500 fb$^{-1}$ to $p$-annihilator WIMPs in terms of the 
values of $\kappa_e$ that can be probed at 3 and 5 $\sigma$ level, as a 
function of the WIMP mass $M_\chi$. (For comparison, a typical value of 
the $\kappa_e$ parameter in the bulk of mSUGRA parameter space is between
0.2 and 0.3.) The kinematic acceptance cuts imposed on the photon are 
$\sin\theta>0.1$, $p_{T,\gamma}>7.5$ GeV. Moreover, the accepted photons have 
to satisfy
\beq
\frac{\sqrt{s}}{2}\left(1-\frac{8M_\chi^2}{s}\right) \leq E_\gamma \leq
\frac{\sqrt{s}}{2}\left(1-\frac{4M_\chi^2}{s}\right).
\eeq{gamma_cut}
The lower cut ensures that the relative motion of the produced WIMPs is 
non-relativistic, $v_\chi^2<1/2$ in the center of mass frame of the WIMP pair. 
The upper cut corresponds to the endpoint of 
the photon spectrum for a given $M_\chi$, and serves to improve the 
signal/background ($S/B$) ratio. Note that the cuts in Eq.~\leqn{gamma_cut}, 
and therefore the dataset used to test the WIMP hypothesis, depend on the 
assumed WIMP mass. 
The values of the signal and background cross sections with 
the cuts specified above for a few representative values of $M_\chi$ are
given in Table~\ref{tab:XS}. The table also lists the signal cross section 
values for the case of $s$-annihilators. It is clear that the sensitivity 
of the proposed search in this case is at best marginal.

\begin{table}
\begin{center}
\begin{tabular}{|c||c|c|c|c|c|c|c|} \hline
   $M_\chi$, GeV & 75 & 100 & 125 & 150 & 175 & 200 & 225  \\ \hline
 $\sigma_{\rm bg}$, fb & 36 & 83 & 202 &  590 & 2030 & 1800 & 1200 \\ \hline
 $\sigma_{\rm sig}$ ($p$-ann.), fb & 1.8 & 3.9 & 8.4 & 21 & 64 & 27 & 4.9 
\\ \hline
 $\sigma_{\rm sig}$ ($s$-ann.), fb & 0.4 & 0.9 & 1.9 & 4.5 & 13 & 8.7 & 3.6  
\\ \hline
\end{tabular}
\caption{Signal and background cross sections at $\sqrt{s}=500$ GeV with 
no polarization and the 
cuts specified in the text, for a few representative values of $M_\chi$. The 
signal cross sections are listed for $p$- and $s$-annihilators with 
$\kappa_e=1$, and scale linearly with $\kappa_e$.}
\label{tab:XS}
\end{center}
\end{table}

The reach of the LC can be further increased by polarizing the beams: while
the background is dominated by the $W$ exchange diagrams which only appear
for left-handed electrons, there is no reason to expect that the WIMP 
couplings have the same asymmetry. For polarized beams, the signal cannot be 
fully characterized by the spin-averaged annihilation fraction $\kappa_e$ 
introduced in Eq.~\leqn{kappa_def}; instead, four independent annihilation 
fractions are needed, corresponding to the four possible $e^+e^-$ helicity 
configurations. To apply Eq.~\leqn{rate} to this case, we make a replacement
\beqa
\kappa_e &\rightarrow& \frac{1}{4}  
(1+P_-)\left[(1+P_+)\kappa(e^R_- e^L_+)+(1-P_+)\kappa(e^R_- e^R_+)
\right]
\CR
& &\hskip-1cm +\frac{1}{4}  
(1-P_-)\left[(1+P_+)\kappa(e^L_- e^L_+)+(1-P_+)\kappa(e^L_- e^R_+)
\right],
\eeqa{replace} 
where $P_\pm$ are the polarizations of the positron and electron beams
($P=0$ corresponds to unpolarized beams, $P_-=1$ to pure right-handed 
electron beam, and $P_+=1$ to pure left-handed positron beam). Ignoring the
$Z$ exchange contribution, the background cross section 
scales as $(1-P_-)(1-P_+)$. For example, let us assume that the WIMP couplings 
to electrons conserve both helicity and parity, $\kappa(e^R_- e^L_+)=
\kappa(e^L_- e^R_+)=2\kappa_e$, $\kappa(e^R_- e^R_+)=\kappa(e^L_- e^L_+)=0$.
Then, for electron polarization $P_-=0.8$ and no positron polarization, the 
$S/B$ ratio is enhanced by a factor of 5, whereas for ($P_-=0.8$, $P_+=0.6$)
the $S/B$ ratio is enhanced by a factor of 18.5 compared to the unpolarized
case.

{\em Discussion ---}
All existing studies discussing the prospects for discovery of WIMPs at 
colliders do so within a specific model, fixing the model parameters to 
reflect the 
observed dark matter abundance. Their interpretation is hindered by the 
large number of theoretical assumptions made about the details of particle 
physics at the TeV scale. 
%For example, such studies often predict much higher rates of WIMP production 
%than what is predicted in this letter, typically due to production 
%of heavier new particles which can subsequently decay into WIMPs. 
For example, supersymmetric models often lead to high rates of events with 
missing $E_T$ at the LHC due to production of strongly-coupled gluinos and
squarks, whose decay chains necessarily involve neutralinos. It is possible,
and indeed likely, that such high rates will be observed. However, this is
by no means guaranteed by the WIMP hypothesis itself. In contrast, the 
rates predicted here are guaranteed (up to the unknown annihilation fraction)
once the WIMP hypothesis is accepted. Therefore, the signatures we have
discussed provide a unique opportunity to directly test the WIMP hypothesis 
at high-energy colliders. 
%Even though searching for them may be challenging experimentally, 
We believe that this direction is well worth pursuing. 

\if
It should be emphasized that while our model-independent approach is only 
valid when the relative velocity of the produced $\chi$'s is non-relativistic
(that is, near the endpoint of the photon spectrum),
%prediction for the 
%reaction $e^+e^-\to2\chi+\gamma$, Eq.~\leqn{rate}, is only valid in the 
%rather narrow kinematic region defined in Eq.~\leqn{validity}, 
experimental searches for anomalous photon+missing energy events can also be 
carried out without this kinematic restriction. The WIMP hypothesis in effect 
guarantees that the rates are non-zero, as long as the relevant annihilation 
fraction does not vanish. While the interpretation of a negative result of 
such a search would be model-dependent, a positive result would provide a 
striking evidence for the validity of the WIMP hypothesis.  
\fi

{\em Conclusions ---} We have developed a robust, model-independent approach 
to computing the rates of WIMP production in particle collisions. Using the
cosmological measurement of the present dark matter abundance, we have 
obtained detailed predictions for the rates of processes relevant for WIMP
searches at particle colliders. Our study highlights the direct connection 
between cosmology and the physics that may be explored by the next generation
of collider-based experiments.  

%%%%%%%%%%%%%%%%%%%%%

{\it Acknowledgments ---} This work is supported by the NSF under grant 
PHY-9513717. K.M. is supported in part by the US DoE under grant 
DE-FG02-97ER41029. We would like to thank the anonymous referee of the earlier 
version of this paper for valuable suggestions.


\begin{thebibliography}{99}
\bibitem{footnote1}
We assume that $\chi$ and $\bar{\chi}$ are identical; the only modification 
required to lift this assumption is the insertion of appropriate factors of
two to undo the identical-particle factors implicit in our formulas.
\end{thebibliography}
\end{document}